# Quantum Electrodynamics and Planck-Scale

## Rainer Collier

Institute of Theoretical Physics, Friedrich-Schiller-Universität Jena

Max-Wien-Platz 1, 07743 Jena, Germany

E-Mail: rainer@dr-collier.de

**Abstract.** This article examines the consequences of the existence of an upper particle momentum limit $p = |\vec{p}| \leq P_*$ in quantum electrodynamics, where $P_* = \sqrt{\hbar c^3/G}$ is the Planck momentum. The method used is Fourier analysis as developed already by Fermi in his fundamental work on the quantum theory of radiation. After determination of the appropriate Hamiltonian, a Schrödinger equation and the associated commutation rules of the field operators are given. At the upper momentum limit mentioned above, the divergent terms occurring in the Hamiltonian (the self-energies of the electrons and the zero-point energy of the electromagnetic field) adopt finite values, which will be stated and compared with each other.

## 1 Introduction

To this day, it is not clear in what way the Planck-scale, i.e. particle energies or momenta in the order of $E_* = M_* c^2$ or $P_* = M_* c$, respectively, can be consistently related to the physics of elementary particles and that of the gravitational field (with $M_* = \sqrt{\hbar c/G} = 2{,}18 \cdot 10^{-5}\, g$ denoting the Planck mass, $\hbar$ the Planck constant, $c$ the velocity of light in vacuo and $G$ Newton's gravitational constant). There exist, though, several suggestions how to solve this problem ad hoc, i.e. without decisive references to experimental experience. Among these approaches there are the string theory, the doubly special relativity theory (DSR), loop quantum gravitation, the introduction of non-commutative geometries, the use of specially deformed Lorentz algebras, as well as several generalized uncertainty principles (GUP) in which the Planck momentum occurs. For some detailed overview articles, see refs. [2], [3] and [4].

To study the consequences of the existence of an upper particle momentum limit for a definite physical object, we had looked at the thermodynamics of the photon gas under this aspect and have proposed corrected Planck radiation laws modified for high photon energies [5]. For the mean spectral energy $\bar{\varepsilon}(p)$ of the energy level $\hbar\omega$ at the temperature $T$, e.g., we obtained the radiation law

$$\bar{\varepsilon}(p) = \frac{\varepsilon(p)}{\exp\left(\dfrac{\varepsilon(p)-\mu}{k_B T}\right) - \left(1 - \dfrac{\varepsilon(p)}{E_*}\right)} \quad , \quad \varepsilon(p) = c\hbar k = \hbar\omega \ . \tag{1.1}$$

For the free energy of the photon gas $F = -PV$ (the thermal equation of state), there followed

$$F(T,V) = g_s \frac{4\pi}{h^3}(k_B T)V \int_0^{P_*} \frac{p^2}{\left(1 - \dfrac{\varepsilon(p)}{E_*}\right)} \ln\left\{1 - \left(1 - \frac{\varepsilon(p)}{E_*}\right)\exp\left[-\left(\frac{\varepsilon(p)-\mu}{k_B T}\right)\right]\right\} dp \ . \tag{1.2}$$

We also calculated the modified caloric equation of state $U = U(T,V)$, the number of particles $N(T,V)$ and the entropy $S(T,V)$. For $P_* = (E_*/c) \to \infty$, all these corrected radiation laws merge into the well-known laws of photon gas thermodynamics.

We also developed a statistical thermodynamics with an upper particle momentum limit for all ideal quantum gases [6].

Because of presumed upper momentum limits, other authors also have derived thermodynamic laws of the photon gas that deviate from Planck's laws, one example being a Planck radiation law modified for high photon energies on the basis of the Lorentz-invariant DSR theory [7].

To study the effects of a maximum particle momentum $p \leq P_* = E_*/c$ also in quantum electrodynamics, we shall follow Fermi's fundamental work on the quantum theory of radiation [1], in which he applies the method of Fourier analysis to the electrodynamic potentials $\vec{a}$ and $\varphi$.

## 2 The classical field equations

Let us start from the Lagrangian density $\mathcal{L}$ of the Maxwell-Dirac field (Gaussian cgs system)

$$\mathcal{L} = c\bar{\psi}\left[\gamma^\mu(p_\mu - \tfrac{e}{c}A_\mu) - m_0 c\right]\psi - \tfrac{1}{4\pi}\cdot\tfrac{1}{4}F_{\mu\nu}F^{\mu\nu} \ , \tag{2.1}$$

with $\psi, \bar{\psi}$ denoting bispinors, $\gamma^\mu$ Dirac spin matrices, $A_\mu$ the four-potential of the electromagnetic field, $e$ the electric charge, $m_0$ the electron rest mass, $F_{\mu\nu} = A_{\nu|\mu} - A_{\mu|\nu}$ the electromagnetic field tensor and $p_\mu = i\hbar\partial_\mu$ the momentum operator. If we select $q_A = (\psi, A_\mu)$ as the independent field variables, we obtain, by means of the Lagrange equations $(\partial\mathcal{L}/\partial q_{A|\nu})_{|\nu} - (\partial\mathcal{L}/\partial q_A) = 0$, the coupled system of field equations

$$\left[\gamma^\mu(p_\mu - \tfrac{e}{c}A_\mu) - m_0 c\right]\psi = 0 \ , \tag{2.2}$$

$$\Box A_\mu = \frac{4\pi}{c} j_\mu \quad , \qquad j_\mu = e c \bar{\psi} \gamma^\mu \psi \quad , \tag{2.3}$$

where $\Box = \partial_\mu \partial^\mu = \partial^2 / \partial (ct)^2 - \vec{\nabla}^2$ and the Lorenz condition

$$A^\mu_{\ |\mu} = 0 \tag{2.4}$$

has been used. In the metric $\eta_{\mu\nu} = diag\,(1,-1,-1,-1)$, the 3- and 4-quantities are related as follows,

$$\gamma^\mu = (\beta, \vec{\gamma}) \quad , \quad p_\mu = i\hbar (\tfrac{1}{c}\partial_t, \vec{\nabla}) \quad , \quad A^\mu = (\varphi, \vec{a}) \quad , \quad j^\mu = (c\rho, \vec{j}) \;, \tag{2.5}$$

$$\vec{\alpha} = \beta \cdot \vec{\gamma} \quad , \quad \vec{\mathcal{E}} = -\tfrac{1}{c}\dot{\vec{a}} - \vec{\nabla}\varphi \quad , \quad \vec{\mathcal{H}} = \vec{\nabla} \wedge \vec{a} \quad , \quad \vec{j} = \rho \dot{\vec{r}} \;. \tag{2.6}$$

The field equation system (2.2) – (2.4), then, has the following 3-dimensional form:

$$i\hbar \frac{\partial \psi}{\partial t} = \left[ c\vec{\alpha} \cdot (\vec{p} - \tfrac{e}{c}\vec{a}) + e\varphi + m_0 c^2 \beta \right] \psi \quad , \tag{2.7}$$

$$\left( \frac{1}{c^2} \frac{\partial^2}{\partial t^2} - \Delta \right) \vec{a} = \frac{4\pi}{c} \vec{j} \quad , \qquad \left( \frac{1}{c^2} \frac{\partial^2}{\partial t^2} - \Delta \right) \varphi = 4\pi \rho \quad , \tag{2.8}$$

$$\frac{1}{c} \frac{\partial \varphi}{\partial t} + \vec{\nabla} \cdot \vec{a} = 0 \;. \tag{2.9}$$

## 3 An approach to the quantum field theory

*3.1 Transition to field operators*

The classical electromagnetic potentials $A_\mu$ become field operators now by being subjected to certain canonical commutation relations, analogously to quantum mechanics. In that context, Fermi developed the electrodynamic potentials $\vec{a}$ and $\varphi$ in Fourier series and regards the solely time-dependent coefficients $\vec{a}_s(t)$, $\varphi_s(t)$ of these series as the proper „coordinates" of the fields,

$$\vec{a} = \sum_s \vec{a}_s(t) \sin \Gamma_s \quad , \tag{3.1}$$

$$\varphi = \sum_s \varphi_s(t) \cos \Gamma_s \;. \tag{3.2}$$

The phase $\Gamma_s$ has the general structure

$$\Gamma_s = \vec{k}_s \cdot \vec{r} + \vartheta_s \quad , \tag{3.3}$$

wherein $\vec{r}$ is the position vector, $\vec{k}_s$ the wave vector and

$$|\vec{k}_s| = k_s = \frac{2\pi}{\lambda_s} = \frac{\omega_s}{c} \qquad (3.4)$$

the wavenumber, $\lambda_s$ the wavelength, $\omega_s$ the angular frequency and $\vartheta_s$ an arbitrary phase shift of the s$^{th}$ partial wave.

Let us now project the vector $\vec{a}_s$ onto three mutually perpendicular unit vectors $\vec{n}_{1s}, \vec{n}_{2s}, \vec{n}_{3s}$, with $\vec{n}_{1s}$ pointing in the propagation direction $\vec{k}_s$ of the s$^{th}$ partial wave,

$$\vec{a}_s = a_{1s}\vec{n}_{1s} + a_{2s}\vec{n}_{2s} + a_{3s}\vec{n}_{3s} \qquad , \qquad \vec{n}_{1s} = \frac{\vec{k}_s}{k_s} \qquad . \qquad (3.5)$$

Now, the field variables of the electromagnetic field are

$$\varphi_s(t), \ a_{1s}(t), \ a_{2s}(t), \ a_{3s}(t) \ . \qquad (3.6)$$

From the differential equations (2.8) and (2.9) we can now determine die time dependencies of the variables (3.6) by integration over the 3-volume $V$. With

$$\int \cos\Gamma_s \cos\Gamma_{s'} dV = \tfrac{1}{2}V\delta_{ss'} \qquad , \qquad g = V/8\pi c^2 \quad , \qquad (3.7)$$

follow the differential equations

$$\ddot{\varphi}_s + \omega_s^2 \varphi_s = \tfrac{1}{g}\int \rho \cos\Gamma_s \, dV \quad , \qquad (3.8)$$

$$\ddot{\vec{a}}_s + \omega_s^2 \vec{a}_s = \tfrac{1}{g}\cdot\tfrac{1}{c}\int \rho \dot{\vec{r}} \sin\Gamma_s \, dV \ . \qquad (3.9)$$

These two differential equations are obviously equivalent to the Maxwell equations if the Lorenz condition (2.9) is also written in terms of the new coordinates (3.6). Considering the orthogonality properties of the base vectors $\vec{n}_{1s}, \vec{n}_{2s}, \vec{n}_{3s}$, we obtain

$$\omega_s a_{1s} + \dot{\varphi}_s = 0 \ , \qquad (3.10)$$

or, with repeated time derivation and (3.8),

$$\omega_s \dot{a}_{1s} - \omega_s^2 \varphi_s + \tfrac{1}{g}\int \rho \cos\Gamma_s dV = 0 \ . \qquad (3.11)$$

*3.2 Transition to point charges*

Following Fermi, we now pass on from a continuous charge distribution to an ensemble of point charges. Let the point charges $e_i$ occupy the places $\vec{r}_i$ so that the integrals in (3.8) to (3.11) can be replaced as follows:

$$\int \rho \cos \Gamma_s \, dV \to \sum_i e_i \cos \Gamma_{si} \quad , \quad \int \rho \dot{\vec{r}} \sin \Gamma_s \, dV \to \sum_i e_i \dot{\vec{r}}_i \sin \Gamma_{si} \quad , \tag{3.12}$$

$$\Gamma_s = \vec{k}_s \cdot \vec{r} + \vartheta_s \to \Gamma_{si} = \vec{k}_s \cdot \vec{r}_i + \vartheta_s \quad . \tag{3.13}$$

The Maxwell equations (with Lorenz condition) of a system of point charges will then adopt the following form

$$\ddot{\varphi}_s + \omega_s^2 \varphi_s = \tfrac{1}{g} \sum_i e_i \cos \Gamma_{si} \quad , \tag{3.14}$$

$$\ddot{\vec{a}}_s + \omega_s^2 \vec{a}_s = \tfrac{1}{g} \cdot \tfrac{1}{c} \sum_i e_i \dot{\vec{r}}_i \sin \Gamma_{si} \quad , \tag{3.15}$$

$$\omega_s a_{1s} + \dot{\varphi}_s = 0 \quad \text{bzw.} \quad \omega_s \dot{a}_{1s} - \omega_s^2 \varphi_s + \tfrac{1}{g} \sum_i e_i \cos \Gamma_{si} = 0 \quad . \tag{3.16}$$

*3.3 Der Hamiltonian operator of the Maxwell-Dirac system*

The Hamiltonian of our Maxwell-Dirac system is formulated as follows:

$$\begin{aligned} \mathbb{H} = \sum_i \sum_s & \left[ c\, \vec{\alpha}_i \cdot \left( \vec{p}_i - \frac{e_i}{c} \vec{a}_s \sin \Gamma_{si} \right) + e_i \varphi_s \cos \Gamma_{si} + m_0 c^2 \beta_i \right] \\ & + \frac{1}{2g} \sum_s \left[ \left( \vec{b}_s^2 + g^2 \omega_s^2 \vec{a}_s^2 \right) - \left( \pi_s^2 + g^2 \omega_s^2 \varphi_s^2 \right) \right] \quad , \quad g = \frac{V}{8\pi c^2} \quad . \end{aligned} \tag{3.17}$$

The form of the term after the double sum in $\mathbb{H}$ we take from the Schrödinger form of the Dirac equation in (2.8) by substituting there the Fourier expansions for $\varphi$ and $\vec{a}$ from (3.1) and (3.2). These terms correspond to the free Dirac field and to the interaction between the Dirac and Maxwell fields. The residual term after the single sum corresponds to the free Maxwell field.

The Maxwell equations with sources can be shown to be contained in the above Hamiltonian (3.17). For this purpose, we use the above $\mathbb{H}$ to write the Hamiltonian canonical equations for the canonically conjugated electromagnetic variables $(\varphi_s, \pi_s)$ and $(\vec{a}_s, \vec{b}_s)$:

$$\dot{\varphi}_s = \frac{\partial \mathbb{H}}{\partial \pi_s} = -\frac{\pi_s}{g} \quad , \quad \dot{\pi}_s = -\frac{\partial \mathbb{H}}{\partial \varphi_s} = g\,\omega_s^2 \varphi_s - \sum_i e_i \cos \Gamma_{si} \quad , \tag{3.18}$$

$$\dot{\vec{a}}_s = \frac{\partial \mathbb{H}}{\partial \vec{b}_s} = +\frac{\vec{b}_s}{g} \quad , \quad \dot{\vec{b}}_s = -\frac{\partial \mathbb{H}}{\partial \vec{a}_s} = -g\,\omega_s^2 \vec{a}_s + \tfrac{1}{c} \sum_i e_i (c\vec{\alpha}_i) \sin \Gamma_{si} \quad . \tag{3.19}$$

By elimination of $\pi_s$ from (3.18), the Maxwell equation (3.14) appears again. If we additionally consider $\dot{\vec{r}}_i = \partial \mathbb{H}/\partial \vec{p}_i = c\vec{\alpha}_i$, the Maxwell equation (3.15) also appears after elimination of $\vec{b}_s$ from (3.19).

Therefore, the fundamental equation of Fermi's quantum theory of an interaction between the electromagnetic field and the electron field reads

$$i\hbar \frac{\partial \Psi}{\partial t} = \mathbb{H}\, \Psi \; . \tag{3.20}$$

Therein, the wave function $\Psi$ has the structure $\Psi = \Psi(\vec{a}_s, \varphi_s, \vec{r}_i, t; \sigma_i)$, with the quantity $\sigma_i$ characterizing the dependence on the electron spin. The differential equation (3.20) have to supplemented by the rules of commutation between the canonical coordinates and the associated momenta,

$$[x_k, p_l] = i\hbar \delta_{kl} \quad , \quad [a_{ks}, b_{ls'}] = i\hbar \delta_{kl}\delta_{ss'} \quad , \quad [\varphi_s, \pi_{s'}] = -i\hbar \delta_{ss'} \; . \tag{3.21}$$

All further combinations of coordinates and momenta commute. Moreover, the Lorenz condition (2.9) or (3.10) and (3.11) needs to be satisfied too. This is effected by splitting off from the $\Psi$–function an exponential factor ($a_{1s}$ and $\varphi_s$ are commutable),

$$\Psi(a_{1s}, a_{2s}, a_{3s}, \varphi_s, \vec{r}_i, t; \sigma_i) = \Phi(a_{2s}, a_{3s}, \vec{r}_i, t; \sigma_i) \cdot \exp[\tfrac{i}{\hbar} \sum_s a_{1s}(g\omega_s \varphi_s - \tfrac{1}{\omega_s}\sum_i e_i \cos\Gamma_{si})] \; , \tag{3.22}$$

which, at the same time, separates the ($a_{1s}, \varphi_s$)-dependence in the Hamiltonian (see chapter 4). We first show the exact fulfilment of the Lorenz condition by satisfying the commutation rules of the electromagnetic field variables in (3.21) by the operators

$$b_{ks} = \frac{\hbar}{i}\frac{\partial}{\partial a_{ks}} \quad , \quad \pi_s = -\frac{\hbar}{i}\frac{\partial}{\partial \varphi_s} \tag{3.23}$$

and applying them to the Lorenz conditions (3.10) and (3.11). With (3.18) and (3.19), these read, in operator form,

$$\left(\pi_s + g\omega_s a_{1s}\right)\Psi = 0 \quad , \quad \left(b_{1s} - g\omega_s \varphi_s + (\tfrac{1}{\omega_s})\sum_i e_i \cos\Gamma_{si}\right)\Psi = 0 \tag{3.24}$$

It is immediately obvious that these two forms of the Lorenz condition are satisfied with the operators (3.23) and the ansatz (3.22) for the $\Psi$–function.

## 4 Discussion of the Hamiltonian operator

Fermi solves the Schrödinger equation (3.20) using the separation ansatz (3.22). In the Schrödinger picture, the operators are known to be time-independent, as a rule. Therefore, the Hamiltonian with $\Psi$ in the form (3.22) is subject to a unitary transformation, so that a new Schrödinger equation results for the new $\Phi$ function:

$$i\hbar \frac{\partial \Phi}{\partial t} = \left(e^{-\Xi}\mathbb{H}\, e^{+\Xi}\right)\Phi = \mathbb{R}\, \Phi \quad , \quad \Xi = \tfrac{i}{\hbar}\sum_s a_{1s}(g\omega_s \varphi_s - \tfrac{1}{\omega_s}\sum_i e_i \cos\Gamma_{si}) \; , \tag{4.1}$$

With the structure of the exponential factor exp($\Xi$) in (4.1), the commutation rules (3.21) satisfied by the approaches (3.23) and with $\dot{\vec{r}}_i = \partial\mathbb{H}/\partial\vec{p} = c\,\vec{\alpha}_i$ taken into account, the new Hamiltonian operator $\mathbb{R}$ indeed has a form that is independent of $a_{1s}$ and $\varphi_s$ and is well interpretable:

$$\mathbb{R} = \sum_i \sum_s \left\{ c\,\vec{\alpha}_i \cdot \left[ \vec{p}_i - \frac{e_i}{c}(a_{2s}\vec{n}_{2s} + a_{3s}\vec{n}_{3s})\sin\Gamma_{si} \right] + m_0 c^2 \beta_i \right.$$
$$\left. + \frac{g}{2}\left[ \dot{a}_{2s}^2 + \dot{a}_{3s}^2 + \omega_s^2(a_{2s}^2 + a_{3s}^2) \right] \right\} + \frac{1}{2g}\sum_s \frac{1}{\omega_s^2}\left( \sum_i e_i \cos\Gamma_{si} \right)^2 . \qquad (4.2)$$

With the terms within the braces, the operator $\mathbb{R}$ describes the radiation problem of an interaction of the electromagnetic field with $i = 1, 2, 3, \cdots$ electrons. The Lorenz convention ensures that field changes occur only transversely to the propagation direction $\vec{n}_{1s} = \vec{k}_s/k_s$ of the respective partial wave.

What about the last term, then, in which no field variables of the electromagnetic field occur? With $\omega_s = ck_s$ and $g = V/8\pi c^2$, it has the form

$$\mathbb{R}_0 = \frac{1}{2g}\sum_i \sum_j e_i e_j \sum_s \frac{\cos\Gamma_{si}\cos\Gamma_{sj}}{k_s^2 c^2} . \qquad (4.3)$$

Let us replace the sum $\sum_s$ by integration over $k$ as follows:

$$\sum_s Z_s \to V\int Z(k)\frac{d^3k}{(2\pi)^3} = V\int Z(k)\frac{4\pi k^2 dk}{8\pi^3} = \frac{V}{2\pi^2}\int Z(k)k^2 dk . \qquad (4.4)$$

Then we get

$$\mathbb{R}_0 = \frac{2}{\pi}\sum_i \sum_j e_i e_j \int_0^\infty \cos\Gamma_i \cos\Gamma_j\, dk , \quad \Gamma_i = \vec{k}\cdot\vec{r}_i + \vartheta , \qquad (4.5)$$

$$\mathbb{R}_0 = \frac{1}{2}\sum_i \sum_j e_i e_j R_{ij} , \quad R_{ij} = \frac{4}{\pi}\int_0^\infty \cos\Gamma_i \cos\Gamma_j\, dk . \qquad (4.6)$$

We calculate $R_{ij}$ by applying the cosine addition theorem and then taking the mean of all directions between $\vec{k}$ and $\vec{r}_i$ and of all phase shifts $\vartheta$. The direction cosine $\mu$ between the vectors $\vec{k}$ and $\vec{r}_i$ fluctuates within a range of $-1 \leq \mu \leq +1$. For $R_{ij}$, we therefore can write

$$R_{ij} = \frac{4}{\pi}\int_0^\infty \frac{1}{2}\int_{-1}^{+1}\cos(k\cdot r_{ij}\cdot\mu)\,d\mu , \quad k = |\vec{k}| , \quad r_{ij} = |\vec{r}_i - \vec{r}_j| , \quad \mu = \cos\tau , \qquad (4.7)$$

with $\tau = \sphericalangle(\vec{k}, \vec{r}_i - \vec{r}_j)$. First we calculate the inner integral, so that there follows

$$R_{ij} = \frac{2}{\pi} \int_0^\infty \frac{\sin(k\, r_{ij})}{k\, r_{ij}} dk = \frac{2}{\pi} \cdot \frac{\pi}{2 r_{ij}} = \frac{1}{r_{ij}} \;. \tag{4.8}$$

Thus, $\mathbb{R}_0$ from (4.6) has the value

$$\mathbb{R}_0 = \frac{1}{2} \sum_i \sum_j \frac{e_i\, e_j}{r_{ij}} \;. \tag{4.9}$$

Hence, the last term $\mathbb{R}_0$ in the Hamiltonian $\mathbb{R}$ (cp. (4.2) and (4.3)) is nothing else than the energy of the Coulomb interaction of all point charges $e_i$ ($i = 1, 2, 3, \cdots$), including the infinitely high self-energy of each individual charge $e_i$ for the case of $i = j$.

## 5 An upper limit to the particle momentum?

In their studies on nonlinear electrodynamics, Born and Infeld [8], followed by Born's introduction of a constantly curved momentum space [9], already introduced a lower length limit or an upper momentum limit, respectively. Recently, Konopka [10] and Moffat [11] also treaded this path.

The assumption of the existence of an upper momentum limit changes the value of the $R_{ij}$ integral (4.8) and, for small spatial distances $r_{ij}$ between two charges $e_i, e_j$, leads to a modified Coulomb law, so that, for $r_{ij} \to 0$, die Coulomb energy remains finite. In the three reports mentioned last, the finiteness of the self-energy of a charge seems to be the result of the introduction of a constant curvature of the momentum space. We can show that the assumption of the existence of an upper particle momentum limit alone (without any more profound explanation by a theoretical model) is sufficient already to guarantee finiteness in the $R_{ij}$ – integral. Let us proceed from (4.6),

$$\mathbb{R}_0 = \frac{1}{2} \sum_i \sum_j e_i e_j R_{ij} \quad , \quad R_{ij} = \frac{4}{\pi} \int_0^{k_*} \cos\Gamma_i \cos\Gamma_j\, dk \;. \tag{5.1}$$

Here, the upper limit $k_*$ is connected with the assumed upper particle momentum limit $P_*$ by

$$P_* = \hbar k_* \quad , \quad k_* = \frac{P_*}{\hbar} = \frac{1}{L_*} \quad , \quad P_* = M_* c \quad , \quad M_* = \sqrt{\frac{\hbar c}{G}} \;, \tag{5.2}$$

with $P_*$ and $M_*$ being the known Planck quantities. The evaluation of $R_{ij}$ in (5.1) yields now

$$R_{ij} = \frac{2}{\pi} \int_0^{k_*} \frac{\sin(k\,r_{ij})}{k\,r_{ij}} dk = \frac{2}{\pi} \cdot \frac{1}{r_{ij}} \int_0^{r_{ij}/L_*} \frac{\sin t}{t} dt = \frac{2}{\pi} \cdot \frac{1}{r_{ij}} \mathrm{Si}\left(\frac{r_{ij}}{L_*}\right) . \tag{5.3}$$

With (5.1) and (5.3), the operator $\mathbb{R}_0$ now has the form

$$\mathbb{R}_0 = \frac{1}{2} \sum_i \sum_j \frac{e_i e_j}{r_{ij}} \cdot \left(\frac{2}{\pi}\right) \mathrm{Si}\left(\frac{r_{ij}}{L_*}\right) . \tag{5.4}$$

The sine integral $\mathrm{Si}(x)$ is known to have the following behaviour at the limits $x \to (0, \infty)$,

$$\lim_{x \to 0} \mathrm{Si}(x) = x - \frac{x^3}{3 \cdot 3!} + \frac{x^5}{5 \cdot 5!} - \cdots \quad , \quad \lim_{x \to \infty} \mathrm{Si}(x) = \frac{\pi}{2} - \frac{\cos x}{x} - \frac{\sin x}{x^3} + \frac{2\cos x}{x^4} + \cdots . \tag{5.5}$$

For great distances $r_{ij} \gg L_*$ ($r_{ij} \to \infty$), $\mathbb{R}_0$ from (5.4) with (5.5) therefore has the value (4.9),

$$\lim_{r_{ij} \to \infty} \mathbb{R}_0 = \frac{1}{2} \sum_i \sum_j \frac{e_i e_j}{r_{ij}} . \tag{5.6}$$

We realize that $\mathbb{R}_0$ is the operator of the electrostatic Coulomb interaction energy of the charges $e_i$ ($i = 1, 2, 3, \cdots$) in the Hamiltonian $\mathbb{R}$ of the radiation problem (cp. (4.2)). For small distances $r_{ij} \ll L_*$ ($r_{ij} \to 0$), there follows from (5.4) with (5.5):

$$\lim_{r_{ij} \to 0} \mathbb{R}_0 = \frac{1}{\pi} \sum_i \frac{e_i^2}{L_*} . \tag{5.7}$$

This is the total self-energy of all charges $e_i$ participating in the radiation process. A single electron with the charge $e$ therefore has the constant self-energy

$$E_0^{(e)} = \frac{1}{\pi} \frac{e^2}{L_*} = \frac{\alpha}{\pi} \frac{\hbar c}{L_*} = \frac{\alpha}{\pi} E_* , \tag{5.8}$$

with the fine structure constant $\alpha = e^2/\hbar c$ and the Planck energy $E_* = M_* c^2$.

Let us now compare the size of the self-energy of the electron $E_0^{(e)}$ with the zero-point energy of the radiation field $E_0^{(\gamma)}$ for the case that also the photon momentum $p = \hbar k \leq P_*$ is limited by the Planck momentum $P_* = M_* c$ (an assumption already made in [5]).

$$E_0^{(\gamma)} = \sum_s \frac{\hbar \omega_s}{2} = \frac{\hbar c}{2} \sum_s k_s = \frac{\hbar c}{2} V \int_0^{k_*} k \frac{d^3 k}{(2\pi)^3} = \frac{\hbar c}{4\pi^2} V \int_0^{k_*} k^3 dk \quad , \quad k_* = \frac{1}{L_*} , \tag{5.9}$$

$$E_0^{(\gamma)} = \frac{\hbar c}{16\pi^2} \frac{V}{L_*^4} = \frac{1}{16\pi^2} \frac{\hbar c}{L_*} \frac{V}{V_*} = \frac{1}{16\pi^2} E_* \frac{V}{V_*} \quad , \quad V_* = L_*^3 \; . \tag{5.10}$$

For the ratio of zero-point energy and self-energy, we get, with (5.8) and (5.10):

$$\frac{E_0^{(\gamma)}}{E_0^{(e)}} = \frac{(1/16\pi^2) E_*}{(\alpha/\pi) E_*} \frac{V}{V_*} = \frac{1}{16\pi\alpha} \frac{V}{V_*} = \frac{137}{16\pi} \frac{V}{V_*} \approx 2.73 \frac{V}{V_*} \; . \tag{5.11}$$

In a Planck volume $V \sim V_* = L_*^3$ (a Planck length cube) there is approximately as much zero-point energy as there is self-energy in an electron.

Moreover, we can also state the maximum number $N$ of the quantum states in a 3-volume $V$:

$$dN = \frac{1}{h^3} d^3 p \, d^3 q \quad , \quad N = \frac{1}{2\pi^2} \frac{V}{\hbar^3} \int_0^{P_*} p^2 dp = \frac{V}{6\pi^2} \frac{P_*^3}{\hbar^3} = \frac{1}{6\pi^2} \frac{V}{V_*} \; . \tag{5.12}$$

This can be understood as a quantization of the 3-dimensional volume, $V = 6\pi^2 \cdot N \cdot V_*$. For the zero-point energy of radiation (5.10), then, we can write $E_0^{(\gamma)} = \frac{3}{8} N E_* \approx \frac{N}{2} E_*$ and find that the zero-point energy of radiation is itself quantized: It solely occurs in a multiple of the Planck energy. The energy amount of the „ground state" ($N = 1$) of this quantized zero-point energy is approximately equal to half the Planck energy.

## 6 Final considerations

In earlier reports, we examined the consequences of the existence of an upper particle momentum limit $p = |\vec{p}| \leq P_* = \hbar k_* = E_*/c$ in the electromagnetic radiation laws [5] and generally in the quantum statistics of ideal gases [6]. With photon gas as an example, we obtained a modified spectral Planck radiation law and, consequently, a generalized thermal equation of state. All these radiation laws corrected for high energies will, for $P_* = (E_*/c) \to \infty$, pass into the familiar laws of the thermodynamics of photon gas.

Supplementary to these new equations of state of photon gas, the present study was aimed at examining the effect of the existence of an upper particle momentum limit $p \leq P_*$ on quantum electrodynamics, and especially on its divergences.

Well suited for the formal establishment of a maximum momentum into quantum electrodynamics is Fermi's introductory treatise [1] on the quantum theory of radiation. First, the Lagrangian and Hamiltonian densities of the Maxwell-Dirac field are written down, and the electrodynamic potentials $\vec{a}$ and $\varphi$ are represented as Fourier series. The Maxwell equations then appear as oscillation differential equations of these potentials with non-zero right-hand sides (3.14) – (3.16), which contain the electric charges. With the associated Hamiltonian $\mathbb{H}$ developed according to Fourier, a Schrödinger equation $i\hbar \dot{\Psi} = \mathbb{H} \Psi$ is set up.

For the canonical coordinates and momenta contained therein, the corresponding commutation rules are formulated. By virtue of a separation ansatz for the wave function $\Psi = \Phi \exp A(a_{1s}, \varphi_s)$, the operator-Lorenz conditions (3.24) are satisfied. For the wave function $\Phi$, then again there applies a Schrödinger equation $i\hbar \dot{\Phi} = \mathbb{R}\Phi$ with $\mathbb{R}$ from (4.2). It is only this Schrödinger-Operator $\mathbb{R}$ that at last describes the pure radiation problem of an interaction of the electromagnetic field with the point charges $e_i \, (i = 1, 2, 3, \cdots)$.

Whereas the Dirac field, Maxwell field and interaction shares appear in the operator $\mathbb{R}$ as usual, there also appears a residual term $\mathbb{R}_0$. Fermi shows that $\mathbb{R}_0$ represents the energy of the Coulomb interaction between the charges $e_i$, which, however, for infinitely small distances between the charges, leads to infinitely great energy amounts. By suggesting a (constantly) curved momentum space, some authors tried to remove this divergence [9], [10], [11]. An earlier reflection on the subject of „structure of the electron" by Weisskopf [12] is still worth reading.

The present article shows that the assumption of the existence of an upper particle momentum limit $p \leq P_*$, on which we proceeded already in refs. [5] and [6], can eliminate some of the divergences in quantum electrodynamics. For example, the value of the electronic self-energy can be determined as $E_0^{(e)} = (1/\pi)(e^2/L_*) = (\alpha/\pi)E_*$, where $E_*$ and $L_*$ are the Planck energy and the Planck length, respectively. If we similarly calculate the zero-point energy of the electromagnetic radiation field in the volume $V$, we obtain $E_0^{(\gamma)} = (1/16\pi^2)E_*(V/V_*)$, with $V_*$ being a cube having the edge length $L_*$. It can be seen that a volume $V \sim V_*$ contains about as much zero-point energy of the radiation field as there is self-energy of one electron. It is further remarkable that this zero-point energy of the electromagnetic radiation itself seems to be quantized. The zero-point energy occurs only as a multiple of the Planck energy $E_*$: $E_0^{(\gamma)} = (3/8)N E_* \approx (N/2)E_*, (N = 1, 2, 3, \cdots)$.

Finally, it should be noted again that these results have been gained due to the assumption of the existence of an invariant upper particle momentum limit $p \leq P_*$, which cannot be explained by the currently accepted theoretical foundations of physics (special theory of relativity and quantum theory). The ongoing experimental and theoretical attempts to change this have been referred to in the introductory chapter of this treatise.

**References**

[1]  E. Fermi, Rev. Mod. Phys. 4 (1932) 87, "Quantum Theory of Radiation".

[2]  G. Amelio-Camelia, Liv. Rev. Rel. 16 (2013) 5, "Quantum space-time phenomenology", arXiv:0806.0339v2 [gr-qc].

[3]  A. N. Tawfik, A. M. Diab, Int. J. Mod. Phys. D23 (2014) 1430025, "Generalized uncertainty principle: approaches and applications", arXiv:1410.0206v2 [gr-qc].

[4]  St. Liberati, Ann. Rev. Nucl. Part. Sci. 59 (2009) 245, "Lorentz Violation: Motivation and new constraints", arXiv:0906.0681v4 [astro-ph.HE].

[5]  R. Collier, arXiv:1109.0427 [gr-qc], "Planck's Radiation Law in the Quantized Universe".

[6]  R. Collier, arXiv:1503.04354 [gr-qc], "Ideal Quantum Gases with Planck Scale Limitations".

[7]  S. Das, S. Pramanik, S. Ghosh, SIGMA 10 (2014)104, "Effects of a Maximal Energy Scale in Thermodynamics for Photon Gas and Construction of Path Integral", arXiv:1411.1839v2 [gr-qc].

[8]  M. Born, L. Infeld, Poc. R. Soc. London A 144 (1934) 425, "Foundations of the New Field Theory".

[9]  M. Born, Proc. R. Soc. London A 165 (1938) 291, "A suggestion for unifying quantum theory and relativity".

[10] T. Konopka, Mod. Phys. Lett. A23 (2008) 319, "A Field Theory Model With a New Lorentz-Invariant Energy Scale", arXiv:hep-th/0601030.

[11] J. W. Moffat, Int. J. Geometric Methods in Mod. Phys. 13(1) (2016) 1650005, "Quantum Gravity Momentum Representation and Maximum Invariant Energy", arXiv:gr-qc/0401117v1

[12] V. F. Weisskopf, Rev. Mod. Phys. 21 (1949) 305, "Recent Developments in the Theory of the Electron".